\documentclass[a4paper,runningheads]{llncs}

\usepackage{butterma}
\idline{This is a post-peer-review, pre-copyedit version of an article published in Lecture Notes in Computer Science Vol. 7167. The final authenticated version is available online at: \url{https://doi.org/10.1007/978-3-642-29645-1_18}.}

\usepackage[utf8]{inputenc}
\usepackage{graphicx}
\usepackage{amsmath}
\usepackage{array}
\usepackage{multirow}
\usepackage{colortbl}
\usepackage{ifthen}

\usepackage[space]{cite}
\usepackage[bookmarks,bookmarksopen]{hyperref}
\hypersetup{
  pdfauthor={Thomas Vogel and Holger Giese},
  pdftitle={Requirements and Assessment of Languages and Frameworks for Adaptation Models},
  pdfsubject={Models in Software Engineering}
}
\newcommand{\RequirementName}[1]{\textbf{\emph{#1}}}
\newcommand{\RequirementNumber}[1]{\textbf{#1}}
\newcounter{LR}
\newcommand{\LanguageRequirement}[2]{
\refstepcounter{LR}
\label{#2}
\noindent\RequirementNumber{LR-\arabic{LR}}\ifthenelse{\equal{#1}{\empty}}{:}{~\RequirementName{#1}:}}
\newcommand{\refLR}[1]{\mbox{LR-\ref{#1}}}
\newcounter{FR}
\newcommand{\FrameworkRequirement}[2]{
\refstepcounter{FR}
\label{#2}
\noindent\RequirementNumber{FR-\arabic{FR}}\ifthenelse{\equal{#1}{\empty}}{:}{~\RequirementName{#1}:}}
\newcommand{\refFR}[1]{\mbox{FR-\ref{#1}}}

\begin{document}
\pagestyle{headings}

\author{Thomas Vogel \and Holger Giese}
\authorrunning{T. Vogel and H. Giese}

\title{Requirements and Assessment of Languages \texorpdfstring{\\}{} and Frameworks for Adaptation Models}
\titlerunning{Adaptation Models: Requirements, Patterns, and Assessment}

\institute{Hasso Plattner Institute at the University of Potsdam, Germany\\
\email{$\{$thomas.vogel,holger.giese$\}$@hpi.uni-potsdam.de}}

\maketitle

\thispagestyle{electronic}

\begin{abstract}
Approaches to self-adaptive software systems use models at runtime to leverage benefits of model-driven engineering (MDE) for providing views on running systems and for engineering feedback loops. Most of these approaches focus on causally connecting runtime models and running systems, and just apply typical MDE techniques, like model transformation, or well-known techniques, like event-condition-action rules, from other fields than MDE to realize a feedback loop. However, elaborating requirements for feedback loop activities for the specific case of runtime models is rather neglected. 

\hspace{2mm} Therefore, we investigate requirements for \emph{Adaptation Models} that specify the analysis, decision-making, and planning of adaptation as part of a feedback loop. In particular, we consider requirements for a modeling language of adaptation models and for a framework as the execution environment of adaptation models. Moreover, we discuss patterns for using adaptation models within the feedback loop regarding the structuring of loop activities and the implications on the requirements for adaptation models. Finally, we assess two existing approaches to adaptation models concerning their fitness for the requirements discussed in this paper.
\end{abstract}

%==============================================================
\section{Introduction}
\label{sec:introduction}

Self-adaptation capabilities are often required for modern software systems to dynamically change the configuration in response to changing environments or goals~\cite{SEfSAS-ROADMAP-2009}. \emph{Models@run.time} are a promising approach for self-adaptive software systems since models may provide appropriate abstractions of a running system and its environment, and benefits of model-driven engineering (MDE) are leveraged to the runtime phases of software systems~\cite{MC.2009.326}.

Most models@run.time efforts to self-adaptive software systems focus on causally connecting models to running systems and just apply typical or well-known techniques from MDE or other fields on top of these models. These techniques are used for engineering a feedback loop that controls self-adaptation by means of \emph{monitoring} and \emph{analyzing} the running system and its environment, and the \emph{planning} and \emph{execution} of changes to the running system~\cite{KephartChess2003}.

For example, the causal connection has been a topic for discussions at the last two workshops on models@run.time~\cite{MRT10-Summary,MRT09report}, or the work of~\cite{MRT10-Song} particularly addresses the causal connection and it just applies MDE techniques, like model transformation, on top to show their technical feasibility. We proposed an approach to use incremental model synchronization techniques to maintain multiple, causally connected runtime models at different abstraction levels, and thereby, we support the monitoring and the execution of adaptations~\cite{VG10,VogelNHGB10}.

While causal connections provide basic support for monitoring and for executing changes, they do not cover the analysis and planning steps of a feedback loop, which decide \emph{if} and \emph{how} the system should be adapted. For these steps, techniques originating from other fields than MDE are used. Most approaches, like~\cite{1537890,georgas-computer09}, employ rule-based mechanisms in some form of event-condition-action rules that exactly specify when and how adaptation should be performed, and thus, the designated target configuration is predefined. In contrast, search-based techniques just prescribe goals that the system should achieve. Triggered by conditions or events and guided by utility functions they try to find the best or at least a suitable target configuration fulfilling these goals~(cf.~\cite{1128711,RamirezMRT2009}).

All these approaches focus on applying such decision-making techniques for the analysis and planning steps, but they do not systematically investigate the requirements for such techniques in conjunction with models@run.time. Eliciting these requirements might help in engineering new or tailored decision-making techniques for the special case of models@run.time approaches to self-adaptive systems. Therefore, we elaborate requirements for such techniques by taking an MDE perspective. The techniques should be specified by models, which we named \emph{Adaptation Models} in an attempt to categorize runtime models~\cite{VogelSG11}. However, the categorization does not cover any requirements for runtime models.

In this paper, which is a revision of~\cite{VG11}, we discuss requirements for adaptation models, in particular requirements for languages to create such models and for frameworks that employ and execute such models within a feedback loop. By language we mean a broad view on metamodels, constraints, and model operations, which are all used to create and apply adaptation models. Moreover, we discuss patterns for using adaptation models within the feedback loop. The patterns and the requirements for adaptation models influence each other, which impacts the design of the feedback loop by providing alternatives for structuring loop activities. Finally, we assess two existing approaches to adaptation models concerning their fitness for the language and framework requirements.

The paper is structured as follows. Section~\ref{sec:related-work} reviews related work, and Section~\ref{sec:adaptation-models} sketches the role of adaptation models in self-adaptive systems. Section~\ref{sec:requirements} discusses the requirements for adaptation models, while Section~\ref{sec:patterns} presents different patterns of employing such models within a feedback loop. Section~\ref{sec:assessment} discusses the assessment of existing approaches to adaptation models with respect to the requirements. The last section concludes the paper and outlines future work.

%==============================================================
\section{Related Work}
\label{sec:related-work}

As already mentioned, most models@run.time approaches to self-adaptive software systems focus on applying techniques for decision-making and do not systematically elaborate on the related requirements~\cite{1537890,1691383,1128711,georgas-computer09,RamirezMRT2009}. A few approaches merely consider the requirement of performance and efficiency for their adaptation mechanisms to evaluate the applicability at runtime~\cite{1128711,RamirezMRT2009}. Likewise, in~\cite{MusicD11} several decision-making mechanisms are discussed in the context of ubiquitous computing applications by means of performance and scalability regarding the size of the managed system and its configuration space. In general, rule-based mechanisms are considered as efficient since they exactly prescribe the whole adaptation, while for search-based approaches performance is critical and often improved by applying heuristics or by reducing the configuration space.

This is also recognized by~\cite{1691383} that attests efficiency and support for early validation as benefits for rule-based approaches. However, they suffer from scalability issues regarding the management and validation of large sets of rules. In contrast, search-based approaches may cope with these scalability issues, but they are not as efficient as rule-based approaches and they provide less support for validation. As a consequence, a combination of rule-based and search-based techniques is proposed in~\cite{1691383} to balance their benefits and drawbacks.

To sum up, if requirements or characteristics of decision-making techniques~are discussed, these discussions are limited to performance, scalability, and support for validation, and they are not done systematically. One exception is the work of Cheng~\cite{OwenCheng2008} who discusses requirements for a self-adaptation language that is focused on specifying typical system administration tasks. However, the requirements do not generally consider self-adaptive software systems and they do not address specifics of models at runtime. Nevertheless, some of the requirements that are discussed in this paper are derived from this work.

%==============================================================
\section{Adaptation Models}
\label{sec:adaptation-models}

\begin{figure}[t]
\centering
\includegraphics[width=.8\columnwidth]{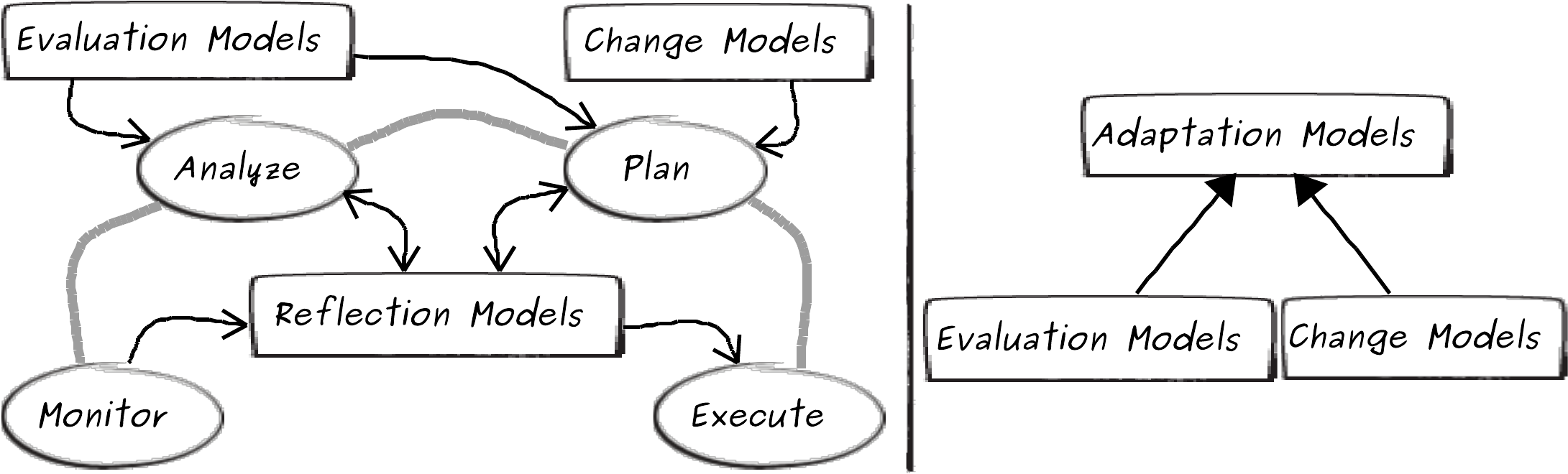}
\vspace{-1mm}
\caption{Feedback Loop and Runtime Models~(cf.~\cite{VogelSG11})}
\label{fig:Loop}
\vspace{-4mm}
\end{figure}

Before discussing requirements for adaptation models, we sketch the role of these models based on a conceptual view on a feedback loop as depicted in Figure~\ref{fig:Loop}. The steps of monitoring and analyzing the system and its environment, and the planning and execution of changes are derived from autonomic computing~\cite{KephartChess2003}, while we discussed the different models and their usage in a feedback loop in~\cite{VogelSG11}. \emph{Reflection Models} describe the running system and its environment, and they are causally connected to the system. According to observations of the system and environment, the monitor updates the reflection models. Reasoning on these models is done by the analyze step to decide whether the system fulfills its goals or not, and thus, whether adaptation is required or not. The reasoning is specified by \emph{Evaluation Models}, which can be constraints that are checked on reflection models. If adaptation is required, the planning step devises a plan defining how the system should be adapted, which is guided by \emph{Change Models} to explore the system's variability or configuration space. Deciding on the designated target configuration is guided by evaluation models to analyze different adaptation options, and the selected option is applied on reflection models. Finally, the execute step performs and effects the adaptations on the running system.

By \emph{Adaptation Models} we generally consider evaluation and change models regardless of the concrete rule-based or search-based techniques that are employed for the analysis and planning steps, and thus, for the decision-making. This view on adaptation models is similar to~\cite{1537890}, which just presents one adaptation model for the specific approach but no general discussion of such models.

%==============================================================
\section{Requirements for Adaptation Models}
\label{sec:requirements}

In this section, we describe requirements for adaptation models to be used in self-adaptive systems to analyze and decide on adaptation needs, and to plan and decide on how to adapt the running system. We assume that the self-adaptive system employs models at runtime, which influence the requirements. At first, we discuss requirements for a modeling language used to create adaptation models. Then, we elaborate the requirements for a framework as the execution environment for adaptation models. Being in the early requirements phase, we take a broad MDE view on the notion of language as combinations of metamodels, constraints, and model operations, which are all used to create and apply models.

Likewise to the common understanding that requirements for real-world applications cannot be completely and definitely specified at the beginning of a software project, we think that the same is true for the requirements discussed here. It is likely that some of these requirements may change, become irrelevant, or new ones emerge when engineering concrete adaptation models for a specific self-adaptive system and domain. Thus, we do not claim that the requirements are complete and finalized with respect to their enumeration and definitions.

%----------------------------------------------------------------
\subsection{Language Requirements for Adaptation Models}
\label{sec:requirements:language}

Language requirements (LR) for adaptation models can be divided into functional and non-functional ones. Functional requirements target the concepts that are either part of adaptation models or that are referenced by adaptation models. These concepts are needed for the analysis, decision-making, and planning. Thus, functional requirements determine the expressiveness of the language. In contrast, non-functional language requirements determine the quality of adaptation models. At first functional, then non-functional requirements are discussed.

% ------------------------------------------------------
% ------------------------------------------------------
\vspace{2mm}
\noindent\textbf{Functional Language Requirements}

% ------------------------------------------------------
\LanguageRequirement{Goals}{LR:Goals} Enabling a self-adaptive system to continuously provide the \mbox{desired} functionality to users or other systems, adaptation models have to know about the current goals of the system. These goals as a functional specification define \emph{what} the system should do, and this information needs to be available in an operationalized form to relate it with the actual behavior of the running system. This is the foundation for adapting the functional behavior of the system.

% ------------------------------------------------------
\LanguageRequirement{Quality Dimensions}{LR:Quality} While~\refLR{LR:Goals} considers \emph{what} the system should do, quality dimensions address \emph{how} the system should provide the functionality in terms of quality of service (QoS). To support QoS-aware adaptations, qualities, like performance or security, must be characterized by adaptation models (cf.~\cite{OwenCheng2008}).

% ------------------------------------------------------
\LanguageRequirement{Preferences}{LR:Preferences} Since multiple quality dimensions (\refLR{LR:Quality}) may be relevant for the managed system, preferences across the dimensions must be expressed to trade-off and balance competing qualities (cf.~\cite{OwenCheng2008}). Likewise, preferences for goals (\refLR{LR:Goals}) are necessary if several valid behavioral alternatives are feasible and not distinguished by the quality dimensions. 

% ------------------------------------------------------
Thus, the language for adaptation models must incorporate the concepts of goals (\refLR{LR:Goals}), qualities (\refLR{LR:Quality}), and preferences (\refLR{LR:Preferences}) in an operationalized form, such that they can be referenced or described and automatically processed by adaptation models. Goals, qualities, and preferences serve as references for the running system as they state what the system should do and how it should~be.

% ------------------------------------------------------
\LanguageRequirement{Access to Reflection Models}{LR:ReflectionModels} Adaptation models must reference and access reflection models to obtain information about the current situation of the running system and environment for analysis and to change the reflection models to effect  adaptations. Thus, a language for adaptation models must be based on the languages of reflection models. 

% ------------------------------------------------------
\LanguageRequirement{Events}{LR:Events} Adaptation models should reference information from events emitted by the monitor step when updating the reflection models due to runtime phenomena of the system. Besides serving as a trigger for starting the decision-making process, events support locating the phenomena in the reflection models (\refLR{LR:ReflectionModels}). Thus, evaluating the system and its environment (\refLR{LR:EvalConditions}) may start right from the point in the reflection models where the phenomena have occurred.

% ------------------------------------------------------
\LanguageRequirement{Evaluation Conditions}{LR:EvalConditions} A language for adaptation models must support the specification of conditions to evaluate the running system and its environment (cf.~\cite{OwenCheng2008}). These conditions relate the goals (\refLR{LR:Goals}), qualities (\refLR{LR:Quality}), and preferences (\refLR{LR:Preferences}) to the actual running system represented by reflection models (\refLR{LR:ReflectionModels}). Therefore, conditions may refer to events notifying about runtime phenomena (\refLR{LR:Events}) as a starting point for evaluation, and they should be able to capture complex structural patterns for evaluating architectural properties.

% ------------------------------------------------------
\LanguageRequirement{Evaluation Results}{LR:EvalResults} Adaptation models must capture the results of computing the evaluation conditions (\refLR{LR:EvalConditions}) because these results identify and decide on adaptation needs especially when the conditions are not met by the system. Adaptation models may annotate and reference the evaluation results in reflection models (\refLR{LR:ReflectionModels}) to locate adaptation needs in the running system.

% ------------------------------------------------------
\LanguageRequirement{Adaptation Options}{LR:AdaptOptions} Adaptation models must capture the variability of the system to know the options for adaptation related to reflection models (\refLR{LR:ReflectionModels}). These options define the configuration space for the system.

% ------------------------------------------------------
\LanguageRequirement{Adaptation Conditions}{LR:AdaptConditions} Adaptation models must consider adaptation conditions since not all adaptation options (\refLR{LR:AdaptOptions}) are feasible in every situation. Thus, conditions should constrain all adaptation options to applicable ones for certain situations (cf.~\cite{OwenCheng2008}). To characterize a situation for an adaptation option, conditions should refer to reflection models (\refLR{LR:ReflectionModels}), events (\refLR{LR:Events}), evaluation results (\refLR{LR:EvalResults}), or other adaptation options. Likewise to such pre-conditions for adaptation options, post-conditions and invariants should be considered.

% ------------------------------------------------------
\LanguageRequirement{Adaptation Costs and Benefits}{LR:AdaptCostsBenefits} Adaptation models should characterize costs and benefits of adaptation options (\refLR{LR:AdaptOptions}) as a basis to select among several possible options in certain situation (cf.~\cite{OwenCheng2008}). Costs should indicate that adaptations are not for free, and benefits should describe the expected effects of adaptation options on the goals (\refLR{LR:Goals}) and qualities (\refLR{LR:Quality}) of the system. By relating costs and benefits to the preferences of the system (\refLR{LR:Preferences}), suitable adaptation options should be selected and applied on the reflection models.

% ------------------------------------------------------
\LanguageRequirement{History of Decisions}{LR:History} Adaptation models should capture a history of decisions, like evaluation results (\refLR{LR:EvalResults}) or applied adaptation options (\refLR{LR:AdaptOptions}), to enable learning mechanisms for improving future decisions.

% ------------------------------------------------------
% ------------------------------------------------------
\vspace{2mm}
\noindent\textbf{Non-functional Language Requirements}

% ------------------------------------------------------
\LanguageRequirement{Modularity, Abstractions and Scalability}{LR:ModularityAbstractions} An adaptation model should be a composition of sub-models rather than a monolithic model to cover all concepts for decision-making. E.g., evaluation conditions (\refLR{LR:EvalConditions}) and adaptation options (\refLR{LR:AdaptOptions}) need to be part of the same sub-model, and even different adaptation options can be specified in different sub-models. Thus, the language should support modular adaptation models. Moreover, the language should enable the modeling at different abstraction levels for two reasons. First, the level depends on the abstraction levels of the employed reflection models (\refLR{LR:ReflectionModels}), and second, lower level adaptation model concepts should be encapsulated and lifted to appropriate higher levels. E.g., several simple adaptation options (\refLR{LR:AdaptOptions}) should be composable to complex adaptation options. Language support for modularity and different abstractions promote scalability of adaptation models.

% ------------------------------------------------------
\LanguageRequirement{Side Effects}{LR:SideEffects} The language should clearly distinguish between con\-cepts that cause side effects on the running system and those that do not. E.g., computing an evaluation condition (\refLR{LR:EvalConditions}) should not affect the running system, while applying an adaptation option (\refLR{LR:AdaptOptions}) finally should. Making the concepts causing side effects explicit is relevant for consistency issues (\refFR{FR:Consistency}).

% ------------------------------------------------------
\LanguageRequirement{Parameters}{LR:Parameters} The language should provide constructs to parameterize adaptation models. Parameters can be used to adjust adaptation models at runtime, like changing the preferences (\refLR{LR:Preferences}) according to varying user needs.

% ------------------------------------------------------
\LanguageRequirement{Formality}{LR:Formality} The language should have a degree of formality that enables on-line and off-line validation or verification of adaptation models, e.g., to detect conflicts or thrashing effects in the adaptation mechanisms.

% ------------------------------------------------------
\LanguageRequirement{Reusability}{LR:Reusability} The core concepts of the language for adaptation models should be independent of the languages used for reflection models in an approach. This leverages the reusability of the language for adaptation models.

% ------------------------------------------------------
\LanguageRequirement{Ease of Use}{LR:EaseOfUse} The design of the language should consider its ease of use because adaptation models are created by software engineers. This influences, among others, the modeling paradigm, the notation, and the tool support. Preferably the language should be based on a declarative modeling paradigm, which is often more convenient and less error-prone than an imperative one. Likewise, appropriate notations and tools are required to support an engineer in creating, validating, or verifying adaptation models.

% ---------------------------------------------------------------
%----------------------------------------------------------------
\subsection{Framework Requirements for Adaptation Models}
\label{sec:requirements:framework}

In the following, we describe framework requirements (FR) for adaptation models. By framework we consider the execution environment of adaptation models, which determines how adaptation models are employed and executed in the feedback loop. Thus, only requirements specific for such a framework are discussed. Typical non-functional requirements for software systems, like reliability or security, are also relevant for adaptation mechanisms, but they are left here.

% ------------------------------------------------------
\FrameworkRequirement{Consistency}{FR:Consistency} The execution of adaptation models should preserve the consistency of reflection models and of the running system. E.g., when adapting a causally connected reflection model, the corresponding set of model changes should be performed atomically and correctly. Thus, the framework should evaluate the invariants, pre- and post-conditions (\refLR{LR:AdaptConditions}) for adaptation options (\refLR{LR:AdaptOptions}) at the model level, before adaptations are executed to the running system.

% ------------------------------------------------------
\FrameworkRequirement{Incrementality}{FR:Incrementality} The framework should leverage incremental techniques to apply or execute adaptation models to promote efficiency. E.g., events (\refLR{LR:Events}) or evaluation results (\refLR{LR:EvalResults}) annotated to reflection models should be used to directly locate starting points for evaluation or adaptation planning, respectively. Or, adaptation options (\refLR{LR:AdaptOptions}) should be incrementally applied on original reflection models rather than on copies. Incrementality could avoid costly operations, like copying or searching potentially large models.

% ------------------------------------------------------
\FrameworkRequirement{Reversibility}{FR:Reversibility} Supporting incremental operations on models (\refFR{FR:Incrementality}), the framework should provide the ability to incrementally reverse performed operations. E.g., the configuration space has to be explored for adaptation planning by creating a path of adaptation options (\refLR{LR:AdaptOptions}) applied on reflection models. Finding a suitable path might require to turn around and to try alternative directions without completely rejecting the whole path. Thus, \emph{do} and \emph{undo} of operations leverages, among others, incremental planning of adaptation.

% ------------------------------------------------------
\FrameworkRequirement{Priorities}{FR:Priorities} The framework should utilize priorities to organize modular adaptation models (\refLR{LR:ModularityAbstractions}) to efficiently and easily identify first entry points for executing or applying adaptation models. E.g., priorities can be assigned to different evaluation conditions (\refLR{LR:EvalConditions}) based on their criticality, and the framework should check the conditions in decreasing order of their criticality.

% ------------------------------------------------------
\FrameworkRequirement{Time Scales}{FR:TimeScales} The framework should simultaneously support different time scales of analysis and adaptation planning. For example, in known and mission-critical situations quick and precisely specified reactions might be necessary (cf. rule-based techniques), while in other situations comprehensive and sophisticated reasoning and planning are feasible (cf. search-based techniques).

% ------------------------------------------------------
\FrameworkRequirement{Flexibility}{FR:Flexibility} The framework should be flexible by allowing adaptation models to be added, removed, and modified at runtime. This supports including learning effects, and it considers the fact that all conceivable adaptation scenarios cannot be anticipated at development-time. Moreover, it is a prerequisite of adaptive or hierarchical control using multiple feedback loops (cf.~\cite{KephartChess2003,VogelSG11}).

%==============================================================
\section{Feedback Loop Patterns for Adaptation Models}
\label{sec:patterns}
In the following, we discuss feedback loop patterns for adaptation models and how the functional language requirements (cf.~Section~\ref{sec:requirements:language}) map to these patterns while considering the framework requirements (cf.~Section~\ref{sec:requirements:framework}). The non-functional language requirements are not further addressed here because they are primarily relevant for designing a language for adaptation models and not for actually applying such models. The patterns differ in the coupling of the analysis and planning steps of a feedback loop, which influences the requirements for adaptation models. Moreover, the adaptation model requirements likely impact the patterns and designs of the loop. Thus, this section provides a preliminary basis for investigating dependencies between requirements and loop patterns.

% ------------------------------------------------------
% ------------------------------------------------------
\subsection{Analysis and Planning -- Decoupled}
\label{sec:patterns:decoupled}

\begin{figure}[tb]
\vspace{1mm}
\begin{minipage}[b]{0.52\linewidth}
\centering
\includegraphics[width=1\linewidth]{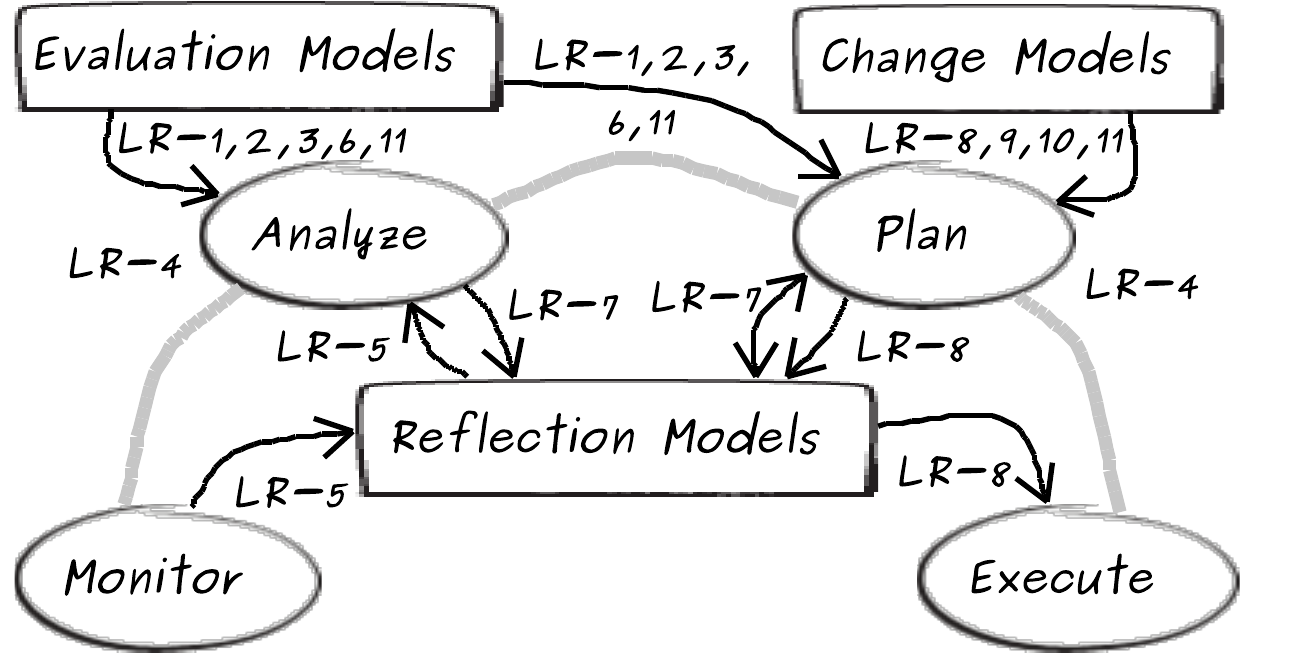}
\caption{Decoupled Analysis and Planning}
\label{fig:LoopReq}
\end{minipage}%
\begin{minipage}[b]{0.48\linewidth}
\centering
\includegraphics[width=1\linewidth]{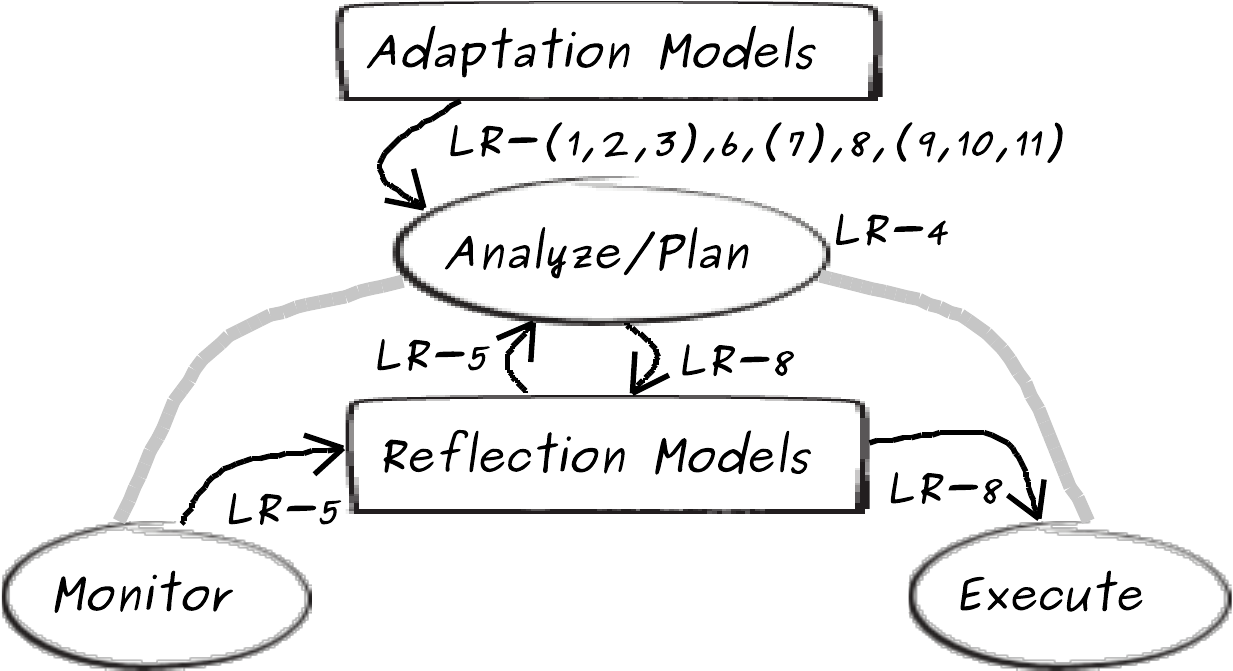}
\caption{Coupled Analysis and Planning}
\label{fig:LoopReq2}
\end{minipage}
\end{figure}

The first pattern of a feedback loop depicted in Figure~\ref{fig:LoopReq} decouples the analysis and planning steps as originally proposed (cf.~Section~\ref{sec:adaptation-models}). The figure highlights functional language requirements (LR) at points where the concepts of the corresponding requirements are relevant. This does not mean that adaptation models must cover all these points, but they must know about the concepts.

In response to events notifying about changes in the running system or environment, the monitor updates the reflection models and annotates the events (\refLR{LR:Events}) to these models. The analyze step uses these events to locate the changes in the reflection models and to start reasoning at these locations. Reasoning is specified by \emph{evaluation models} defining evaluation conditions (\refLR{LR:EvalConditions}) that relate the goals (\refLR{LR:Goals}), qualities (\refLR{LR:Quality}), and preferences (\refLR{LR:Preferences}) to the characteristics of the running system. These characteristics are obtained by accessing reflection models (\refLR{LR:ReflectionModels}). Analysis is performed by evaluating the conditions and probably enhanced by consulting past analyses (\refLR{LR:History}). This produces analysis results (\refLR{LR:EvalResults}) that are annotated to the reflection models to indicate adaptation needs. The planning step uses these results (\refLR{LR:EvalResults}) attached to reflection models (\refLR{LR:ReflectionModels}) to devise a plan for adaptation. Planning is based on \emph{change models} specifying adaptation options (\refLR{LR:AdaptOptions}) and their conditions  (\refLR{LR:AdaptConditions}), costs, and benefits (\refLR{LR:AdaptCostsBenefits}). This information and probably plans devised in the past (\refLR{LR:History}) are used to find suitable adaptation options to create potential target configurations by applying these options on reflection models. These reflection models that prescribe alternative target configurations are analyzed with the help of evaluation models to select the best configuration among them. In contrast to the analyze step that uses evaluation models to reason about the current configuration (descriptive reflection models), the planning step uses them to analyze potential target configurations (prescriptive reflection models). Finally, the execute step enacts the selected adaptation options (\refLR{LR:AdaptOptions}) to the running system.

This pattern is similar to the generic idea of search-based approaches since planning is done by exploring adaptation options (\refLR{LR:AdaptOptions},\,\ref{LR:AdaptConditions},\,\ref{LR:AdaptCostsBenefits}) that are evaluated (\refLR{LR:EvalConditions},\,\ref{LR:EvalResults},\,\ref{LR:History}) for their fitness for the preferenced system goals (LR-\ref{LR:Goals},\,\ref{LR:Quality},\,\ref{LR:Preferences}) based on the current situation of the system and environment (\refLR{LR:ReflectionModels}). Explicitly covering all language requirements for adaptation models, this pattern rather targets comprehensive and sophisticated analysis and planning steps working at longer time scales (\refFR{FR:TimeScales}), while efficiency concerns could be tackled by incrementality.

This pattern leverages incrementality (\refFR{FR:Incrementality}) since the coordination between different steps of the loop is based on events, analysis results, and applied adaptation options, which directly point to location in reflection models for starting analysis, planning, or executing changes. Moreover, analysis and planning steps may incrementally interleave. Based on first analysis results that are produced by evaluation conditions with highest priorities (\refFR{FR:Priorities}), a planning process might start before the whole system and environment have been completely analyzed. However, incrementality requires the reversibility of performed operations (\refFR{FR:Reversibility}) to ensure consistency of reflection models (\refFR{FR:Consistency}), e.g., when alternative adaptation options are tested on-line on reflection models and finally discarded.

% ------------------------------------------------------
% ------------------------------------------------------
\subsection{Analysis and Planning -- Coupled}
\label{sec:patterns:coupled}

In contrast to decoupling the analyze and planning steps, they can be closely integrated into one step, which is sketched in Figure~\ref{fig:LoopReq2}. Based on events (\refLR{LR:Events}), the integrated analyze/plan step computes evaluation conditions (\refLR{LR:EvalConditions}) that are directly mapped to adaptation options (\refLR{LR:AdaptOptions}). If a condition is met, the corresponding adaptation options are applied on the reflection models and finally executed to the running system. Access to reflection models (\refLR{LR:ReflectionModels}) is realized by the analyze/plan step as a link between adaptation and reflection models.

In Figure~\ref{fig:LoopReq2}, the language requirements written in brackets are not explicitly covered by adaptation models because this pattern precisely specifies the adaptation by directly relating evaluation conditions to the application of adaptation options. This relation implicitly covers some of the requirements listed in brackets. E.g., it is assumed that the applied adaptation options effect the system in a way that fulfills the desired goals, qualities, and preferences (LR-\ref{LR:Goals},\,\ref{LR:Quality},\,\ref{LR:Preferences}).

Considering the events and the mapping of evaluation conditions to \mbox{adaptation} options, this pattern is similar to rule-based approaches using event-conditions-action rules. Covering the whole decision-making process and integrating analysis and planning into one step, adaptation models as depicted in Figure~\ref{fig:LoopReq2} cannot~be clearly differentiated into evaluation and change models.

Thus, this pattern targets adaptation mechanisms requiring quick reactions to runtime phenomena by enabling adaptation at rather short time scales (\refFR{FR:TimeScales}). Moreover, efficiency is improved by incrementality (\refFR{FR:Incrementality}) and priorities (\refFR{FR:Priorities}). The steps may incrementally coordinate each other through locating events and applied adaptation options in reflection models for analysis/planning and executing changes to the system. Priorities may be used to order evaluation conditions for quickly identifying critical situations that need urgent reactions, while conditions for non-critical situations can be evaluated without strict time constraints.

The consistency requirement (\refFR{FR:Consistency}) is not explicitly covered because it is assumed that the mapping of conditions to adaptation options preserves consistency by design of such rule-based mechanisms. Since these mechanisms strictly prescribe the adaptation, there need not to be any options left that have to be decided at runtime. This reduces the need for reversible operations (\refFR{FR:Reversibility}).

% ------------------------------------------------------
% ------------------------------------------------------
\subsection{Discussion}

Regarding the two different feedback loop patterns and their effects on adaptation models, we can make two observations. First, it might be necessary to combine both patterns in a self-adaptive system if simultaneous support for different time scales (\refFR{FR:TimeScales}) is required or if a specific self-adaptive system requires both flavors of rule-based and search-based decision-making mechanisms. Second, we think that these two patterns span a range of several other patterns. By explicitly covering more and more language requirements, the adaptation models get more elaborate, and we may move stepwise from the coupled pattern (cf. Section~\ref{sec:patterns:coupled}) toward the decoupled one (cf. Section~\ref{sec:patterns:decoupled}). Which pattern and adaptation models suit best depends on the concrete self-adaptive system, especially on the system's domain requirements. Finally, the requirement of flexibility (\refFR{FR:Flexibility}) has not been discussed for the two patterns. However, it is relevant for both of them since it is usually not possible to anticipate all adaptation scenarios at development-time, which requires adjusting adaptation models at runtime.

%==============================================================
\section{Assessment of Approaches to Adaptation Models}
\label{sec:assessment}

In this section, we assess two approaches to adaptation models, namely \emph{Stitch}~\cite{OwenCheng2008} and \emph{Story Diagrams}~\cite{GHS09}, concerning their support for the requirements presented in this paper. After sketching both approaches, the assessment is discussed.

Stitch is a self-adaptation language developed in the context of \emph{Rainbow}~\cite{OwenCheng2008}, which is a framework for self-adaptive systems based on architecture description language~(ADL) techniques, in particular the \emph{Acme} ADL. The focus of Stitch is to capture routine system administration tasks as explicit adaptation strategies consisting of condition-action pairs. The conditions expressed in a first-order predicate logic are evaluated on an ADL-based reflection model (cf.~Section~\ref{sec:adaptation-models}) describing the running system. If a condition is met indicating a need for adaptation, the actions associated with this condition are analyzed based on utility preferences and the most promising action is directly executed to the system.

Story Diagrams~\cite{GHS09}, originally introduced in~\cite{Fischer+1998}, are a general purpose graph transformation approach. They extend activity diagrams from the \emph{Unified Modeling Language}~(UML) by specifying each activity using a graph transformation rule, called \emph{Story Pattern}. Thus, a Story Diagram defines the control flow between multiple Story Patterns. Story Diagrams and Patterns are specified on one or more user-defined metamodels and they work on corresponding instances of these metamodels. In the context of adaptation, they work on reflection models (cf.~Section~\ref{sec:adaptation-models}). A Story Pattern specifies a pattern that has to be matched in the reflection model. If a match has been found, the side effects of the rule -- if any are specified -- are executed by changing the model. Moreover, patterns can be extended with constraints specified in the \emph{Object Constraint Language}~(OCL) to allow more sophisticated conditions. Following MDE principles, Story Diagrams leverage the usage of MDE techniques, like OCL. Moreover, Story Diagrams themselves conform to a metamodel, which enables an interpreter to directly execute them and which in the end supports higher order transformations. 

% ------------------------------------------------------
% ------------------------------------------------------
\subsection{Assessment of Stitch and Story Diagrams} 

Having outlined both approaches to adaptation models, we analyze them with respect to their support for the requirements discussed in Section~\ref{sec:requirements}. Table~\ref{table:requirements-stitch-SD} lists all requirements and shows the degree of support by both approaches. It has to be noted that both languages make use of other languages, primarily languages for reflection models, to specify an adaptation. E.g., an adaptation model may specify a condition, like \emph{component.rt\,$>$\,MAX\_RT}, to identify whether the response time of a running component exceeds a threshold. While the response times are provided by reflection models (\emph{component.rt} is part of the reflection model language), the adaptation model just defines the threshold and references the reflection model. Thus, Stitch and Story Diagram models are not self-contained.

\begin{table}[t]
\caption{Requirements (Req.), Stitch, and Story Diagrams (SD): '--' denotes no support, '$M$' medium support, and '$F$' full support for the requirements.}
\label{table:requirements-stitch-SD}
\begin{center}
\scalebox{.8}{
\begin{tabular}{|>{\centering}m{12mm}|>{\centering}m{12mm}|>{\centering}m{12mm}|>{\centering}m{.1mm}|>{\centering}m{12mm}|>{\centering}m{12mm}|>{\centering}m{12mm}|>{\centering}m{.1mm}|>{\centering}m{12mm}|>{\centering}m{12mm}|>{\centering}m{12mm}|}
	\hline
\multicolumn{11}{|l|}{{\cellcolor[gray]{0.8}\textbf{Functional Language Requirements}}} \tabularnewline \hline
    Req. 								& Stitch 	& SD	&{\cellcolor[gray]{0.8}}&
	Req. 								& Stitch 	& SD 	&{\cellcolor[gray]{0.8}}&
	Req.								& Stitch	& SD	\tabularnewline \hline
	\refLR{LR:Goals}					& --		& $F$	&{\cellcolor[gray]{0.8}}&
	\refLR{LR:Events}					& $M$		& $F$	&{\cellcolor[gray]{0.8}}&
	\refLR{LR:AdaptConditions}			& $F$		& $F$	\tabularnewline \hline
	\refLR{LR:Quality}					& $F$		& $F$	&{\cellcolor[gray]{0.8}}&
	\refLR{LR:EvalConditions}			& $F$		& $F$	&{\cellcolor[gray]{0.8}}&
	\refLR{LR:AdaptCostsBenefits}		& $F$		& $F$	\tabularnewline \hline
	\refLR{LR:Preferences}				& $F$		& $F$	&{\cellcolor[gray]{0.8}}&
	\refLR{LR:EvalResults}				& --		& $F$	&{\cellcolor[gray]{0.8}}&
	\refLR{LR:History}					& $M$		& $F$	\tabularnewline \hline
	\refLR{LR:ReflectionModels}			& $M$		& $F$	&{\cellcolor[gray]{0.8}}&
	\refLR{LR:AdaptOptions}				& $F$		& $F$	&{\cellcolor[gray]{0.8}}&
	\multicolumn{3}{c|}{}									\tabularnewline \hline \hline
\multicolumn{11}{|l|}{{\cellcolor[gray]{0.8}\textbf{Non-functional Language Requirements}}} \tabularnewline \hline
    Req. 								& Stitch 	& SD	&{\cellcolor[gray]{0.8}}&
	Req. 								& Stitch 	& SD 	&{\cellcolor[gray]{0.8}}&
	Req.								& Stitch	& SD	\tabularnewline \hline
	\refLR{LR:ModularityAbstractions}	& $M$		& $F$	&{\cellcolor[gray]{0.8}}&
	\refLR{LR:Parameters}				& $M$		& $F$	&{\cellcolor[gray]{0.8}}&
	\refLR{LR:Reusability}				& $M$		& $M$	\tabularnewline \hline
	\refLR{LR:SideEffects}				& --		& $M$	&{\cellcolor[gray]{0.8}}&
	\refLR{LR:Formality}				& --		& $M$	&{\cellcolor[gray]{0.8}}&
	\refLR{LR:EaseOfUse}				& $M$		& $F$	\tabularnewline \hline \hline
\multicolumn{11}{|l|}{{\cellcolor[gray]{0.8}\textbf{Framework Requirements}}} \tabularnewline \hline
    Req. 								& Stitch 	& SD	&{\cellcolor[gray]{0.8}}&
	Req. 								& Stitch 	& SD 	&{\cellcolor[gray]{0.8}}&
	Req.								& Stitch	& SD	\tabularnewline \hline
	\refFR{FR:Consistency} 				& $M$		& $F$	&{\cellcolor[gray]{0.8}}&
	\refFR{FR:Reversibility}			& --		& $M$	&{\cellcolor[gray]{0.8}}& 
 	\refFR{FR:TimeScales}				& --		& $F$	\tabularnewline \hline
	\refFR{FR:Incrementality}			& --		& $M$	&{\cellcolor[gray]{0.8}}& 
 	\refFR{FR:Priorities} 				& --		& $F$	&{\cellcolor[gray]{0.8}}&
	\refFR{FR:Flexibility}				& --		& $F$	\tabularnewline	\hline
\end{tabular}
}
\end{center}
\vspace{-6mm}
\end{table}

Concerning functional language requirements, Stitch focuses on QoS-aware adaptation and thus, it provides full support for quality dimensions (\refLR{LR:Quality}) and preferences across these dimensions (\refLR{LR:Preferences}). Functional goals (\refLR{LR:Goals}) are not considered. Story Diagrams may provide full support for goals, qualities, and preferences (LR-\ref{LR:Goals},\,\ref{LR:Quality},\,\ref{LR:Preferences}) as they work on reflection models and use OCL, which is similar to Stitch. To cover goals, Story Diagrams may even use an explicit goal model describing the designated functionality in addition to reflection models.

Access to reflection models (\refLR{LR:ReflectionModels}) is supported by both languages though Stitch strategies just have read access for analysis, but they do not perform changes on the model before executing them to the system. This might be helpful for model-based planning or testing of adaptation and thus, Stitch provides medium support. Story Diagrams and especially Story Patterns explicitly read and write (change) reflection models, such that we attest them full support here.

While Stitch uses events (\refLR{LR:Events}) only as triggers for the adaptation process to compute all evaluation conditions, the Story Diagram interpreter uses as well the information contained in change events to locate points in reflection models where evaluation conditions as Story Patterns should be checked. Thus, event information is used to filter the conditions irrelevant for these locations, which improves efficiency. This motivates the medium resp. full support for events.

Evaluation conditions (\refLR{LR:EvalConditions}) and adaptation conditions (\refLR{LR:AdaptConditions}) are supported by both approaches based on the integration of some form of first-order predicate logic (\emph{Acme} predicates in case of Stitch, and OCL for Story Diagrams). However, Story Patterns provide additional means to specify structural conditions by means of patterns containing structured model elements to be checked.

Stitch does not explicitly capture evaluation results (\refLR{LR:EvalResults}) as they are used in a transient way to select adaptation strategies. In contrast, Story Diagrams may provide full support by employing Story Patterns just for analysis purposes. Thus, the pattern to be matched specifies the evaluation condition and the corresponding side effects compute results that are annotated to the reflection models.

Both approaches provide full support for adaptation options (\refLR{LR:AdaptOptions}) and adaptation costs and benefits (\refLR{LR:AdaptCostsBenefits}). Likewise to Stitch that uses multiple strategies and that defines cost-benefit attributes of each adaptation step, multiple Story Diagrams or multiple Story Patterns enriched with such attributes are feasible for specifying and selecting appropriate adaptations to be performed.

Finally, since Stitch does not clearly separate the analysis and planning steps, it just maintains a history (\refLR{LR:History}) for the applied adaptation strategies, but not for the analysis results as they are not explicitly captured. In contrast to this medium support, Story Diagrams or Patterns explicitly addressing evaluation results as well as the applied adaptation options may keep a history of both.

Concerning non-functional language requirements, Stitch partially supports modularity, abstractions, and scalability (\refLR{LR:ModularityAbstractions}) by the strategy, tactic, and \mbox{operator} concepts. Operators are system-level commands that are bundled in tactics to describe an adaptation step, and a strategy orchestrates multiple of these steps. Thus, Stitch is limited to these three levels of abstraction. In contrast, besides Story Diagrams and Story Patterns already provide an initial abstraction, Story Diagrams can be nested in other Story Diagrams without any restrictions.

Stitch does not distinguish between concepts causing side effects or not~(\refLR{LR:SideEffects}) as the strategies are considered as inherently causing effects on the running system. In contrast, Story Patterns can be statically analyzed whether they potentially cause side effects as well as they can be annotated to make it explicit.

Parameters (\refLR{LR:Parameters}) are supported by both. While Stitch seems to be restricted on parameters of basic data types, Story Diagrams and Patterns may have arbitrary parameters including references to objects of user-defined classes.

While the Stitch language is not based on a formal foundation (\refLR{LR:Formality}), Story Patterns built upon the graph transformation theory. This enables support for formal validation and verification, which is, however, impeded if OCL is used.

Regarding reusability (\refLR{LR:Reusability}) both languages, Stitch and Story Diagrams, are similar as they are independent of the languages used for the reflection models. However, the concrete adaptation models created with Stitch or Story Diagrams use and reference concepts of the reflection model languages. Thus, the concrete adaptation models are coupled to the types of reflection models.

Stitch is basically an imperative scripting language and its tool support requires improvement~\cite{OwenCheng2008}. For Story Diagrams and Patterns, the declarative notion of graph transformations and graphical editors assist an engineer in modeling and validating adaptation models. Therefore, checks for syntactical well-formedness, an interpreter, and a visual debugger are provided. This causes the different ratings of both approaches concerning the ease of use requirement (\refLR{LR:EaseOfUse}).

Finally, the support for the framework requirements is discussed. Since Stitch's adaptation strategies that have been selected to tackle the current adaptation need are directly executed to the running system, only limited support for consistency (\refFR{FR:Consistency}) is provided. Consistency is only addressed by observing intermediate effects of the executed adaptation on the running system but not beforehand at the model level. The other framework requirements are not covered by Stitch.

Since Story Diagrams completely work on reflection models, consistency can be continuously checked at the model level. Incrementality (\refFR{FR:Incrementality}) is supported for single Story Patterns, and reversibility (\refFR{FR:Reversibility}) for typical model changes by tracking primitive operations performed on the model. Both requirements are hard to satisfy if OCL is used. Incremental evaluation of OCL statements is often not possible and for the case of side effects the inverse statements might not be detectable. Prioritizing (\refFR{FR:Priorities}) Story Diagrams and Patterns to be executed is supported, and it is required for different time scales (\refFR{FR:TimeScales}). E.g., a Story Pattern defining the whole adaptation for urgent situations must have a higher priority to be executed than other Patterns that jointly define a sophisticated adaptation by separating analysis and planning steps. Since Story Diagrams and Patterns are interpreted at runtime, they can replaced or modified on-line, e.g., by higher order transformations. This satisfies the flexibility requirement (\refFR{FR:Flexibility}).
\vfill
% ------------------------------------------------------
% ------------------------------------------------------
\subsection{Discussion of the Assessment}

The conducted assessment of Stitch and Story Diagrams concerning the requirements presented in this paper is constrained by two aspects. First, the analysis of Stitch is solely based on literature~\cite{OwenCheng2008}. Second, the analysis of Story Diagrams for their fitness to specify adaptation models is based on our experience with Story Diagrams and thus, it has been done from a conceptual point of view.

The fact that Stitch does not support several requirements does not \mbox{necessarily} reveal design flaws, but it is rather motivated by the specific setting. On the one hand and as argued in~\cite{OwenCheng2008}, Stitch focuses on system administration tasks that could be tackled well by rule-based approaches and it limits runtime \mbox{reasoning on} computing utilities to select one among multiple applicable adaptation \mbox{strategies}. Thus, Stitch does not aim to support search-based mechanisms. On the other hand, Stitch is not based on MDE principles and it does not take into account specifics of models@run.time. However, using an ADL model at runtime, Stitch is still related to the research field for which the requirements are relevant.

Concerning the different feedback loop patterns discussed in Section~\ref{sec:patterns}, Stitch targets the pattern that couples the analysis and planning steps, which is similar to rule-based approaches. However, since Stitch supports utility-based analysis and selection of competing adaptation strategies (rules), it shares characteristics with the decoupled pattern. This motivates the need for explicitly capturing, e.g., qualities and preferences (LR-\ref{LR:Quality},\,\ref{LR:Preferences}), which need not to be the case for pure rule-based approaches that do not allow competing or even conflicting rules.

Finally, Story Diagrams seem to be a promising language for adaptation models because they follow MDE principles, they are directly interpretable and flexible, and they seamlessly integrate with any user-defined metamodel for reflection models. However, Story Diagrams are a general-purpose language and not specifically tailored for adaptation models. This might be a drawback, but we think that concepts specific for adaptation models (like goals (\refLR{LR:Goals})) can be covered by a different language (like for goal models), which can be integrated with Story Diagrams. Which of the different feedback loop patterns (cf.~Section~\ref{sec:patterns}) can be covered by Story Diagrams requires further investigations or even experiences from employing Story Diagrams in a concrete self-adaptive system.

%==============================================================
\section{Conclusion and Future Work}
\label{sec:conclusion}

In this paper, we have elaborated the requirements for adaptation models that specify the decision-making process in self-adaptive software systems using models@run.time. In particular, requirements for a modeling language incorporating metamodels, constraints, and model operations for creating and applying adaptation models have been discussed, as well as requirements for a framework that executes adaptation models. Moreover, we discussed patterns of a self-adaptive system's feedback loop with respect to the requirements for adaptation models. Additionally, we have assessed two existing languages and frameworks, namely Stitch and Story Diagrams, concerning their fitness for the requirements.

As future work, we want to elaborate the conducted assessment to confirm evidence for the relevance and completeness of the requirements. Moreover, we plan to use Story Diagrams for specifying adaptation models in our approach~\cite{VG10,VogelNHGB10} to close the feedback loop. Additionally, this can be seen as an evaluation of the promising results that Story Diagrams achieved in the assessment.

%==============================================================

%==============================================================
\end{document}